\documentclass{article}

\usepackage[utf8]{inputenc} 
\usepackage{longtable}

\usepackage{graphicx}

\usepackage{url}
\usepackage[cmex10]{amsmath}
\usepackage{mathrsfs}
\usepackage{amssymb}
\usepackage{amssymb}
\usepackage{amsmath}
\usepackage{latexsym}
\usepackage{amsthm}

\newtheorem{dfn}{Definition}[section]

\usepackage{url}
\usepackage{hyperref}
\newcommand{\footremember}[2]{%
    \footnote{#2}
    \newcounter{#1}
    \setcounter{#1}{\value{footnote}}%
}
\newcommand{\footrecall}[1]{%
    \footnotemark[\value{#1}]%
} 

\begin{document}



\title{Towards context in large scale biomedical knowledge graphs}

\author{%
	Jens Dörpinghaus\footnote{jens.doerpinghaus@scai.fraunhofer.de} \footremember{fh}{Fraunhofer Institute for Algorithms and Scientific Computing SCAI, Schloss Birlinghoven, Sankt Augustin, Germany}%
	\and Andreas Stefan\footrecall{fh}%
	\and Bruce Schultz\footrecall{fh}%
 	\and Marc Jacobs\footrecall{fh}%
	}




\date{\today}

\maketitle

\begin{abstract} 
 Contextual information is widely considered for NLP and knowledge discovery in life sciences since it highly influences the exact meaning of natural language. The scientific challenge is not only to extract such context data, but also to store this data for further query and discovery approaches. Here, we propose a multiple step knowledge graph approach using labeled property graphs based on polyglot persistence systems to utilize context data for context mining, graph queries, knowledge discovery and extraction. 
We introduce the graph-theoretic foundation for a general context concept within semantic networks and show a proof-of-concept based on biomedical literature and text mining. Our test system contains a knowledge graph derived from the entirety of PubMed and SCAIView data and is enriched with text mining data and domain specific language data using BEL. Here, context is a more general concept than annotations. This dense graph has more than 71M nodes and 850M relationships. We discuss the impact of this novel approach with 27 real world use cases represented by graph queries. 
\end{abstract}


%



\section{Background}

The amount of available and stored data is constantly increasing in many areas in the course of digitalization. The increasing amount of data represents a great challenge for storage and requires the development of new storage technologies. At the same time, with more available data and different storage technologies, new applications based on the data are of great interest. Large data collections are used for \emph{data mining} and \emph{knowledge discovery} to answer new and complex questions more efficiently. For this purpose, data is often stored in non-relational databases, and while there are many types available, one of the more interesting and promising types are \emph{knowledge graphs}. In this database structure, the entities of a domain are stored as nodes in a graph while connections between these entities are represented by edges. This allows for visualization and analysis of networks between the data in order to discover new applications. 

Current systems use RDF (Resource Description Framework) Triple Stores, systems that inherently have some serious limitations especially when compared to a labeled property graph. For example nodes and edges have no internal structure which does not allow complex queries like subgraph matchings or traversals and it is not possible to uniquely identify instances of relationships which have the same type, see \cite{desai2018issues}. Several approaches have been made to create RDF knowledge graphs, for example Bio2RDF (see \cite{dumontier2014bio2rdf} and \cite{callahan2013bio2rdf}, reviewed by \cite{li2014research} or \cite{natsiavas2015exploring}). For our generalized concept of context, we require labeled property graph structures. 

\emph{Context} is a widely discussed topic in text mining and knowledge extraction since it is an important factor in determining the correct semantic sense of unstructured text. In \cite{aggarwal2012introduction}, Nenkova and McKeown discuss the influence of context on text summarization. Ambiguity is an issue for both common language words and those in scientific context. The challenge in this field is not only to extract such context data, but also to be able to store this data for further natural language processing (NLP), querying and discovery approaches. Here, we propose a multiple step knowledge graph based approach to utilize context data for biological resarch and knowledge expression based on our results published in \cite{dorpinghaus2019knowledge}. We present a proof of concept using biomedical literature and present an outlook on additional improvements which can be implemented in the next generation of knowledge extraction e.g. training approaches from artificial intelligence and machine learning. 

Knowledge graphs have been shown to play an important role in recent knowledge mining and discovery.  A \emph{knowledge graph} (sometimes also called a \emph{semantic network}) is a systematic way to connect information and data to knowledge on a more abstract level compared to language graphs. This type of data structure has many advantages in terms of searching within biomedical data and serves as a vital tool capable of generating novel ideas. Another important attribute when generating knowledge is context and therefore connecting knowledge graphs using contextual information can further enhance data anlysis and hypothesis generation.

As a basis for this work, we generated a knowledge graph that initially contains publication metadata from \emph{PubMed}\footnote{\url{https://www.ncbi.nlm.nih.gov/pubmed}} which has more than 30 million documents at its disposal, including biomedical publications. In subsequent steps, the knowledge graph was expanded to include \emph{BEL} (Biological Expression Language) relations and named entities obtained from text mining using JProMiner (see \cite{Hanisch2005}) and stored in SCAIView\footnote{\url{https://www.scaiview.com/}} as well as ontologies or terminologies like \emph{MeSH}. This results in a large amount of data for the graph with a very high number of nodes and edges. Saving and managing such a graph poses challenges due to the horizontal scalability of graph databases, therefore, it is to be expected that search queries on the graph have a long runtime. 
This paper presents a polyglot persistence approach to tackle this challenge using Neo4j\footnote{\url{https://neo4j.com/}}, a graph database with a native graph storage. 

Here, we use a general definition of context data assuming that each information entity can also be contextual information for other entities, for example a document can also serve as context for other documents (e.g. by citing or referring to the other publication). An author is both metainformation for a document, but also itself context (by other publications, affiliations, co-author networks, ...). Other data is more obviously purely context: named entities, topic maps, keywords, etc. extracted with text mining from documents. However, relations extracted from a text document may stand for themselves, occurring in multiple documents and still valuable without the original textual information. 

\begin{figure*}[t]
	\begin{center}
		\includegraphics[width=0.8\textwidth]{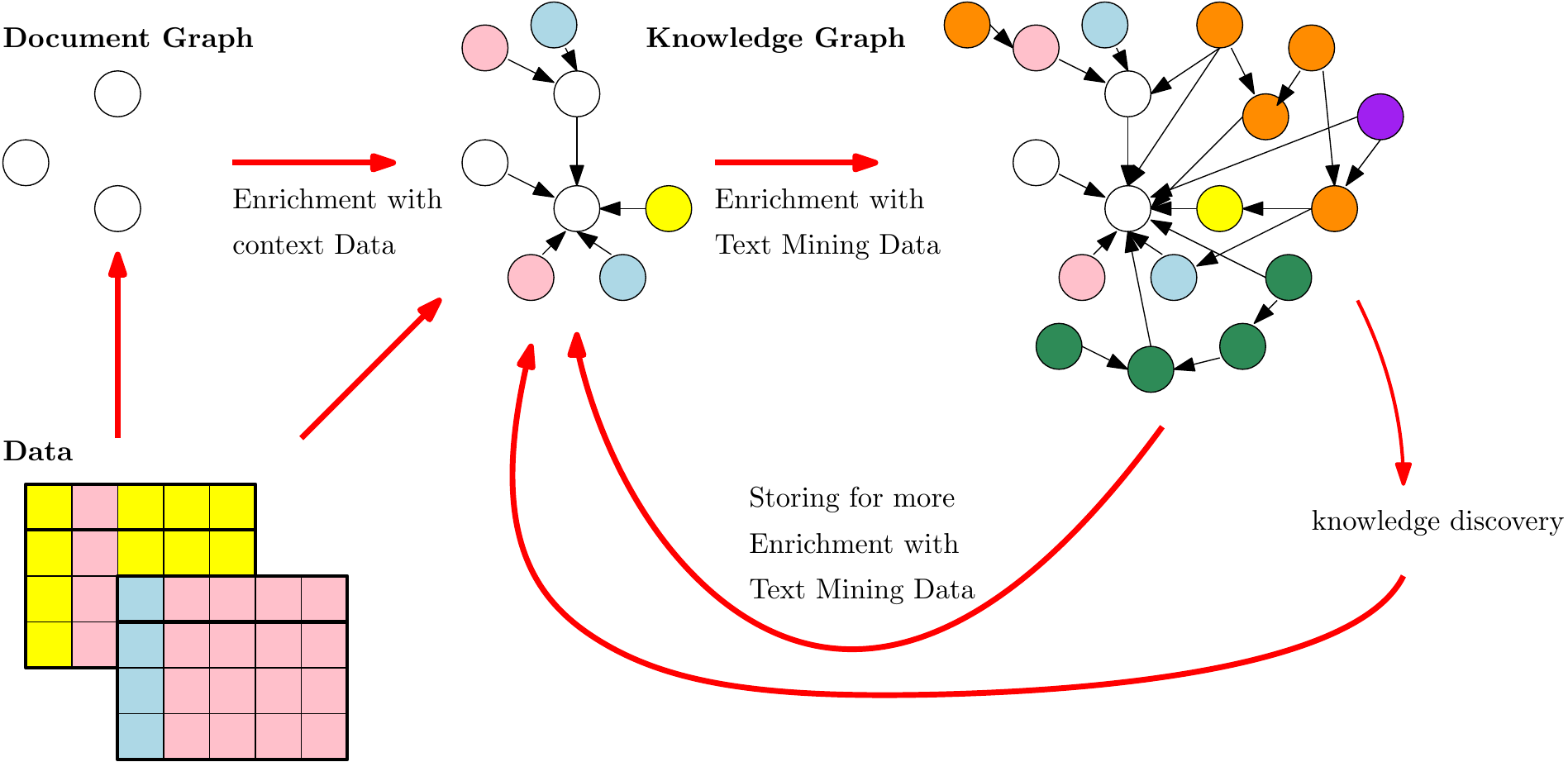}
		\caption{
			Proposed workflow to extend a knowledge graph. First starting with a document graph, the basic meta information like authors, keywords etc. are added. This can be used as a basis for text mining which can be used to extend the graph again, for example  named entity recognition (NER) may use keywords as context. Topic detection may also benefit from already assigned keywords, journals or author information. The graph can also be extended by knowledge discovery processes, for example finding parameters of a clinical trial, progression within electronic health records, etc. In any case new context information are added to the initial graph and improve the input of further algorithms. 
		}
		\label{fig.steps}
	\end{center}
\end{figure*}

To start, we begin with a simple document graph and, in the first step, we added context metainformation (see Figure \ref{fig.steps}). This leads to an initial knowledge graph which can be used for preliminary context-based text mining approaches. In doing so, additional context data is be added to the knowledge graph, such as entities or concepts from ontologies or relations extracted from the analyzed text. The resulting knowledge graph can be used as starting basis for more detailed text mining approaches which utilize the novel context data. These steps can be repeated several times to further enrich the graph. 

In fact, using a graph structure to house data has several additional advantages for knowledge extraction: biological and medical researchers, for example, are interested in exploring the mechanisms of living organisms and gaining a better understanding of underlying fundamental biological processes of life. 
Systems biology approaches, such as integrative knowledge graphs, are important to decipher the mechanism of a disease by considering the system as a whole, which is also known as the holistic approach. To this end, disease modeling and pathway databases both play an important role. Knowledge graphs built using BEL are widely applied in biomedical domain to convert unstructured textual knowledge into a computable form. The BEL statements that form knowledge graphs are semantic triples that consist of concepts, functions and relationships \cite{fluck}. 
In addition, several databases and ontologies can implicitly form a knowledge graph. For example Gene Ontology, see \cite{GO} or DrugBank, see \cite{wishart2017drugbank} or \cite{khan2019consensus} cover a large amount of relations and references to which reference other fields. 

There are still several crucial issues to consider when converting literature to knowledge such as evaluating the quality and completeness of such networks. Furthermore, in order to generate new knowledge, context of concepts in a knowledge graph must be considered. 

To start, we first present a preliminary overview about information theory and management. Afterwards, we will introduce and discuss the novel approach of managing and mining contextual data of knowledge graphs. Finally, we will give a detailed list of issues that need to be addressed and show the results from evaluating real use cases.

\subsection{Preliminaries}

A \emph{knowledge graph} is a systematic way to connect information and data to knowledge. It is thus a crucial concept on the way to generate knowledge and wisdom, to search within data, information and knowledge. As described above, context is the most important topic to generate knowledge or even wisdom. Thus, connecting knowledge graphs with context is a crucial feature. 

\begin{dfn}(Knowledge Graph) 
We define a knowledge graph as graph $G=(E,R)$ with entities $e\in E=\{E_1,...,E_n\}$ coming from a formal structure $E_i$ like ontologies. 
\end{dfn}

The relations $r\in R$ can be ontology relations, thus in general we can say every ontology $E_i$ which is part of the data model is a subgraph of $G$ indicating $O\subseteq G$. In addition, we allow inter-ontology relations between two nodes $e_1, e_2$ with $e_1 \in E_1$, $e_2 \in E_2$ and $O_1 \neq E_2$. In more general terms, we define $R=\{R_1,...,R_n\}$ as a list of either inter-ontology or inner-ontology relations. Both $E$ as well as $R$ are finite discrete spaces.

Every entity $e\in E$ may have some additional metainformation which needs to be defined with respect to the application of the knowledge graph. For instance, there may be several node sets (some ontologies, some document spaces (patents, research data, ...), author sets, journal sets, ...) $E_{1},...,E_{n}$ so that $E_{i}\subset {E}$ and ${E} = \cup_{i=1,...,n} E_{i}$. The same holds for ${R}$ when several context relations come together such as "is cited by", "has annotation", "has author", "is published in", etc. 

\begin{dfn} (Context)
We define context $C$ as a set with context subsets $C=\{c_{1},...,c_{m}\}$. This is a finite, discrete set. Every node $v\in G$ and every edge $r\in R$ may have one or more contexts $c\in C$ denoted by $con(v)\subset G$ or $con(r)\subset G$. 
\end{dfn}

It is also possible to set $con(v)=\emptyset$.  Thus we have a mapping $con:E\cup R\rightarrow \mathcal{P}(C)$. If we use a quite general approach towards context, we may set $C=E$. Therefore, every inter-ontology relation defines context of two entities, but also the relations within an ontology can be seen as context, 

With the neighborhood $N(E_i)$ every node set  $E_{i} \in \{E_{1},...,E_{n}\}$ induces a subgraph $G[E_{i}]\subset G$: 

\begin{dfn} (Extended Context Subgraph, Graph Embeddings)
With $G^c[E_i]=G[E_i]\cup N(E_i)$ we denote the extended context subgraph which also contains the neighbors of each node in $G$, which is context of that node. 
\end{dfn}

\begin{figure*}[t]
	\begin{center}
		\includegraphics[width=0.8\textwidth]{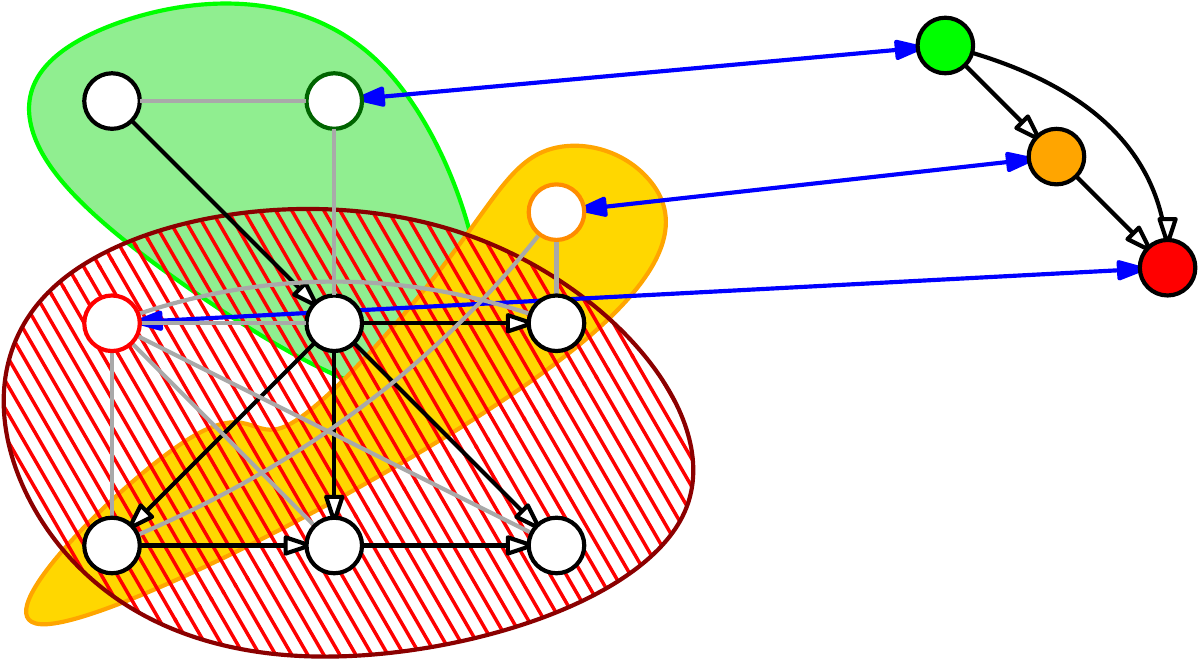}
		\caption{
			Illustration of a \emph{knowledge graph with context} (left). Context is illustrated by colored nodes (green, red, orange) connected to nodes. The colored areas describe the extended context subgraph or context embedding of these contexts.  At the right the corresponding \emph{context metagraph} is described. Every context in the knowledge graphs refers to a node in the metagraph. The references in the original knowledge graph are illustrated by a blue edge. The edges within the metagraph describe if in the original graph an edge from one context to the next exist. 
		}
		\label{fig.metagraph}
	\end{center}
\end{figure*}

For a graph drawing perspective, if $G^c[E_i]$ defines a proper surface, we can think about a graph embedding of another subgraph $G^c[E_j]$ on $G^c[E_i]$. This concept was introduced in \cite{dorpinghaus2019semantic}. Here, semantic knowledge graph embeddings were displayed between different layers. Every layer (for example: molecular layer, document layer, mechanism layer) corresponds to another context defining new contexts on other layers. See Figure \ref{fig.metagraph} for an illustration.

\begin{dfn} (Context Metagraph)
We can create the metagraph $M=(C,R')$ of these contexts. Each context is identified by a node in $M$. If there is a connection in $G$ between two contexts, we add an edge $(c_{1},c_{2})\in R'$. This means if 
$\exists (v_{1},v_{2}) \in R: \; c_{1}\in con(v_{1}),\,c_{2}\in con(v_{2})$ $\Rightarrow$ $(c_{1},c_{2})\in R'$ or  
$\exists  (v_{1},v_{2}) \in R: \; c_{1}\in con( (v_{1},v_{2}) ), \,c_{2}\in con(v_{2})$ $\Rightarrow$ $(c_{1},c_{2})\in R'$ or  
$\exists  (v_{1},v_{2}) \in R: \; c_{1}\in con( v_{1} ), \,c_{2}\in con( (v_{1},v_{2}) )$  $\Rightarrow$ $(c_{1},c_{2})\in R'$.
\end{dfn}

Adding edges between the knowledge graph $G$ or a subgraph $G'=(E',R')\subseteq G=(E,R)$ and the metagraph $M$ in $G\cup M$ will lead to a novel graph. 
This can be either seen as inverse mapping $con^{-1}(G')$ or as the hypergraph $\mathcal{H}(G')=(X,\hat{E})$ given by
\[X=E'\cup G^c[E_i]\]
\[\hat{E}=\{ \{e_i, e \forall e\in N(e_i)\} \forall e_i \in X\}\]

\begin{figure*}[t]
	\begin{center}
		\includegraphics[width=0.6\textwidth]{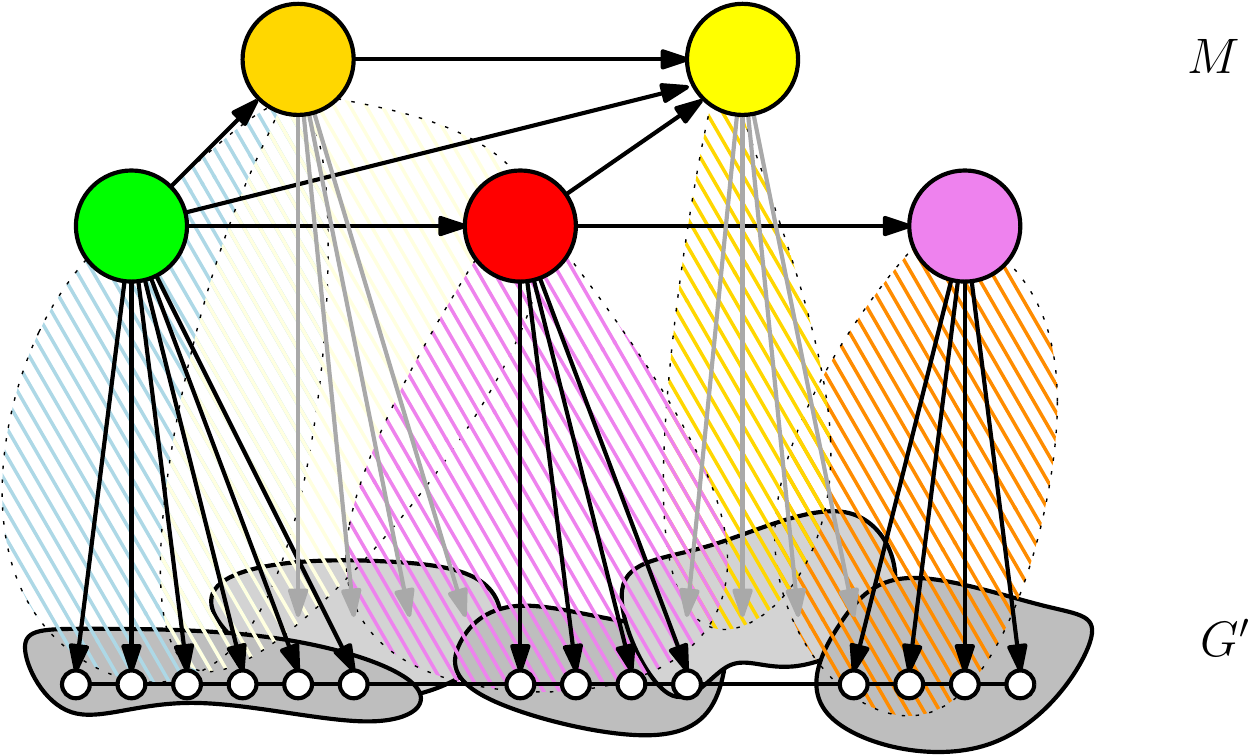}
		\caption{
			This figure describes the hypergraph $\mathcal{H}(G')=(X,\hat{E})$ between the context metagraph $M$ and the original knowledge graph $G$ or a subgraph $G'\subset G$. This graph is sorted by contexts. The hyperedges, illustrated by sets and indicated by non-hyperedges, connect nodes with context, but also nodes with the same context.
		}
		\label{fig.metagraph2}
	\end{center}
\end{figure*}

This graph can be seen as an extension of the original knowledge graph $G'$ where contexts connect not only to the initial nodes, but also every two nodes in $G'$ are connected by a hyperedge if they share the same context as shown in Figure \ref{fig.metagraph2}.

If $C=E$, this will lead to new edges in $G$ thus enriching the original graph. This step should be performed after every additional extension of graph $G$.

We denote this hypergraph $H$ on a knowledge graph $G$ and a metagraph $M$ with $H_{G|M}$. We can add multiple metagraphs $M_{1}$ and $M_{2}$ which is denoted by $H_{G|M_{1},M_{2}}$.

The resulting graph can thus be seen as an enrichment of the original knowledge graph $G$ with contexts. It can be used to answer several research questions and to find graph-theoretic formulations of research questions. 

If the mapping $con$ is well defined for the domain set, then Graph $H$ can be generated in polynomial time. Since this is generally not the case, this step usually contains data or text mining task to generate other contexts from free texts or knowledge graph entities. With respect to the notation described in \cite{dorpinghaus2018question} this problem $p$ can be formulated as
\begin{equation} \label{eq:1}
	p=\mathbb{D}|R|\mathbf{f}:\mathbb{D}\rightarrow\mathbb{X}|err|\emptyset
\end{equation}
Here, the domain set $\mathbb{D}$ is explicitly given by $\mathbb{D}=G$ or -- if additional full-texts $\hat{D}$ supporting the knowledge Graph $G$ exist -- $\mathbb{D}=\{G,\hat{D}\}$, which in our case is the domain subset $R=\mathbb{D}$. Therefore, we need to find a description function $f:\mathbb{D}\rightarrow\mathbb{X}$ with a description set $\mathbb{X}=C$ which holds all contexts. To find relevant contexts, we also need to measure the error as defined by $err:\mathbb{D}\rightarrow[0,1]$.

Several research questions must be considered. First, what metainformation can be used to generate context for a new metagraph? Several promising candidates include authors, citations, affiliation, journal, MeSH-terms and other keywords since they are all available in most databases. We also need to discuss text mining results such as NER, relationship mining etc. Having more general data including study data, genomics, images, etc. we might also consider side effects; disease labels, population labels (male; female; age; social class; etc.). Figure \ref{fig.steps} shows a proof of concept for a less complex text mining metadata approach which describes the process of starting with a simple document graph that can be extended with more context data derived from text mining. We discuss this in more detail in the next section.

The second research question addresses the application of this novel approach for both biomedical research as well as text classification and clustering, NLP and knowledge discovery, with a focus on Artificial Intelligence (AI). How can we use the context metagraph to answer biomedical questions? What can we learn from connections between contexts and how do they look like in the knowledge graph? How can we use efficient graph queries utilizing context? It may also be useful to filter paths in the knowledge graph according to a given context or to generate novel visualizations. A possible question might be to learn about mechanisms linked to co-morbidities or mechanisms being contextualized by drug information. The meta-graph may also contain information about cause-and-effect relationships in the knowledge graph that are “valid” in a biomedical sense under certain conditions as well as contextualization based on demographic information or polypharmacy information. We will discuss several use cases in the last section of this paper.

\subsection{Method}

\subsubsection{Technical setup}
We illustrate the following methods with example runs on PubMed and PMC data. Both sources are already included in the SCAIView NLP-pipeline. PubMed contains 30 million abstracts from biomedical literature, while PMC houses nearly 4 million full-text articles.

First and foremost, the knowledge graph must be stored and accessed by the software in an efficient manner. To this end, a software component was written to integrate the knowledge graph into our SCAIView microservice architecture, see \cite{scaiview}. 
This integration also ensures that the knowledge graph is constantly updated with preprocessed data.  
The software component also provides an API to execute several queries on the knowledge graph and is capable of returning the result in JSON Graph Format\footnote{\url{http://jsongraphformat.info/}} which can be easily displayed by many frontend frameworks.  

Our software component was written in Java using Spring Boot\footnote{\url{http://spring.io/projects/spring-boot}} and Spring Data\footnote{\url{https://spring.io/projects/spring-data}} to be able to access the database backend in an abstract way and ensure the exchangeability of the database technology. 
The database backend in our case is the graph database Neo4j\footnote{\url{https://neo4j.com/}}. 
Neo4j supports the possibility to perform an initial bulk import, allowing us to import the massive knowledge graph in one easy step. The bulk import tool of Neo4j requires that the input data is in the CSV file format. To this end, we designed a software component that exports the data derived from SCAIView as CSV files. 

Storing a large knowledge graph from PubMed, such as the one presented here, in a single database is not a simple task, and we expected the execution of our graph queries to be very slow due to the size of the knowledge graph. To speed up the run times of the queries, we decided to implement an approach that divides the graph using polyglot persistence. 
Polyglot persistence is defined as combining heterogenous data storing technologies into a single application. Instead of storing all of the data in one database, we chose to store different parts of the data in different database technologies. The benefit of polyglot persistence is that each database technology has different strengths and the application can take advantage of them all. 
 
\begin{figure*}[t]
	\begin{center}
		\includegraphics[width=0.8\textwidth]{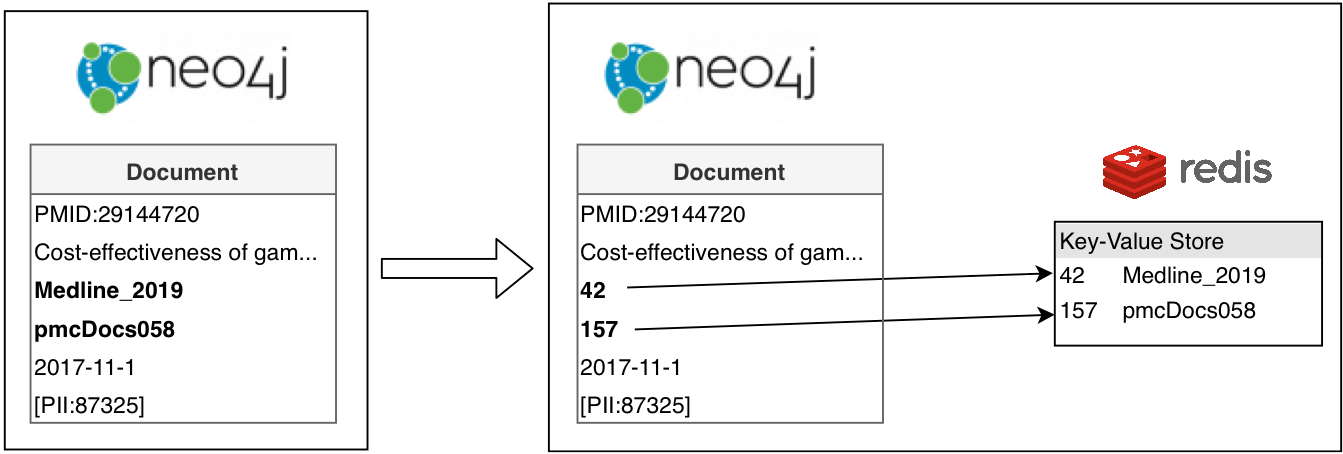}
		\caption{
Example of a stored document node in Neo4j. On the left side a PubMed document is stored with all of its attributes. Using polyglot persistence we see on the right side the same document storing integer encoding for two attributes in Neo4j. The encoding of the used attributes is stored in the key-value database Redis.   
		}
		\label{fig.polyglot}
	\end{center}
\end{figure*}

In Neo4j, the graph structure is stored separately from the properties of nodes and edges. This organization structure makes traversing the knowledge graph easier, however, storing and accessing string attributes takes longer than integer attributes because of this property \cite{Webber2015}.
To take advantage of this characteristic of Neo4j, we designed a storing system that encodes the string attributes of the graph as integers using polyglot persistence. By encoding and storing these attributes in key-value databases, we reduced the data size of the knowledge graph and were able to speed up the property access of Neo4j. 
Figure \ref{fig.polyglot} provides an illustration of the designed polyglot persistence system.

In two iterations, we selected suitable attributes of all node types thus leading to three systems: the original one using only Neo4j (called \textit{Full}) and two polyglot persistence systems (called \textit{Poly1} and \textit{Poly2}). \emph{Full} stores all data directly in Neo4j. \emph{Poly1} stores a few information in another redis database while \emph{Poly2} combines multiple redis databases and the Neo4j graph database. 

\begin{figure*}[t]
	\begin{center}
		\includegraphics[width=0.8\textwidth]{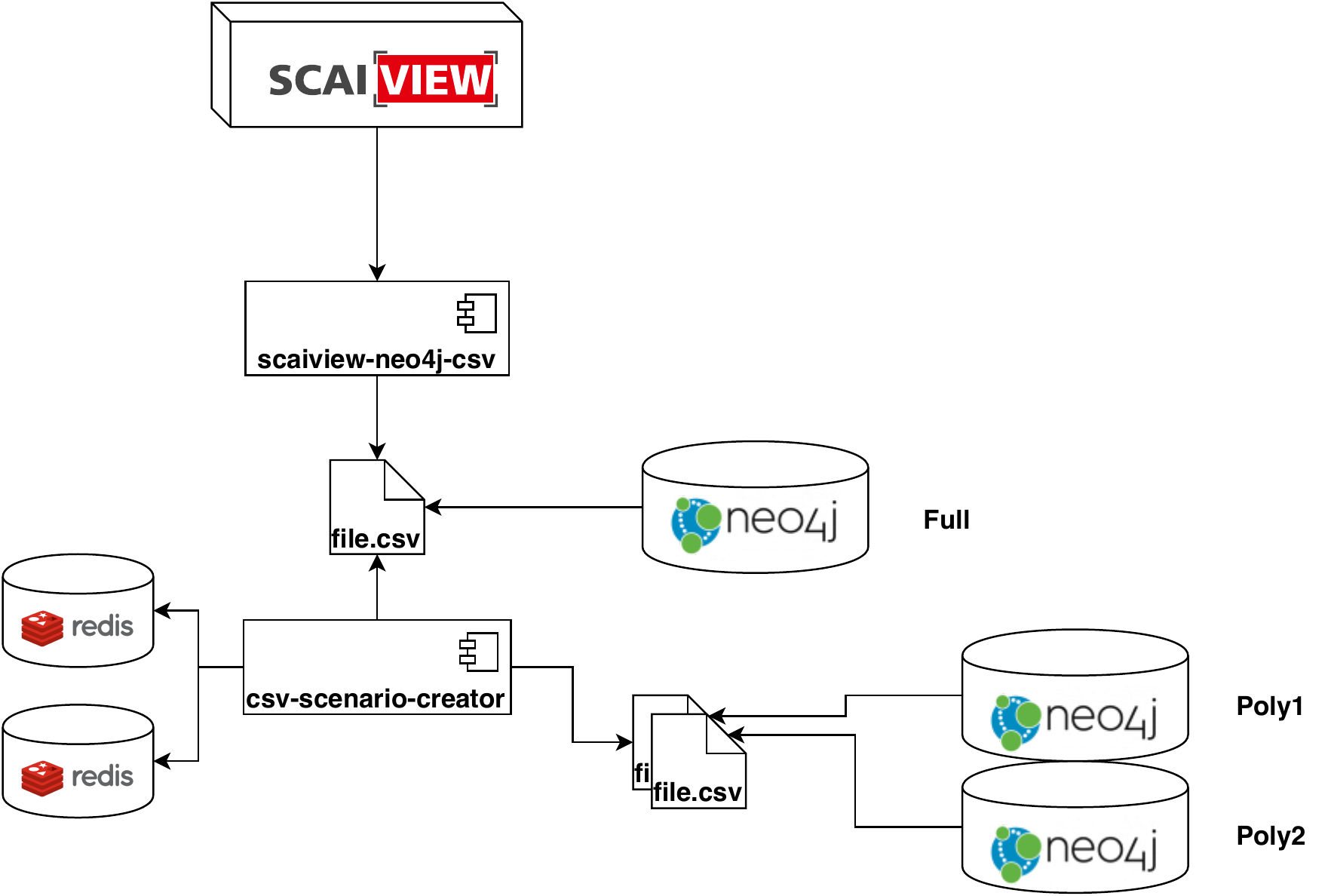}
		\caption{
The software component \texttt{scaiview-neo4j-csv} creates CSV files for the bulk import in Neo4j from SCAIView data. The created files are used as input for the system called \textit{Full}. The second software component \texttt{cdv-scenario-creator} uses the CSV files, runs the encoding of the selected string attributes and created CSV import files for \textit{Poly1} and \textit{Poly2}. 
		}
		\label{fig.Kontextsicht}
	\end{center}
\end{figure*}

We implemented another software component to execute the data preprocessing step for \textit{Poly1} and \textit{Poly2}. It uses the created CSV input files of \textit{Full} to run the data encoding in key-value databases and generates CSV input files for the Neo4j graph databases of the polyglot persistence systems. The whole process is illustrated in Figure \ref{fig.Kontextsicht}.

To compare the execution runtime of queries on all three systems \textit{Full}, \textit{Poly1} and \textit{Poly2}, we collected 27 real word graph queries using the given knowledge graph. The results of the query runtimes are discussed in Section \ref{sec.results}.

\subsubsection{Creating a document and context graph with basic context extraction}

The first step in creating a document and context graph with basic context extraction is to define the entity sets $E_{1},...,E_{n}$ and their relations. 
The articles and abstracts from PubMed and PMC already contain a lot of contextual data. We may define $E_{Document}$ as the document set containing nodes, with each one representing one document. Furthermore, we may add a set $E_{Source}=\{\text{PubMed, PMC}\}$ as the source of a document. Thus, each document can be interpreted as contextual data of a particular data source. 

All meta data are stored in new node sets. $E_{Author}$ stores the set of authors and $E_{Affiliation}$ stores their affiliation, which is again considered context for the authors. Another relevant piece of contextual information is the publisher, in our case $E_{Journal}$. PubMed has several classifications for $E_{Journal}$ including: Books and Documents, Case Reports, Classical Article, Clinical Study, Clinical Trial, Journal Article, and Review. We store this classification in $E_{PublicationType}$.

\begin{figure}[t]
	\begin{center}
		\includegraphics[width=0.45\textwidth]{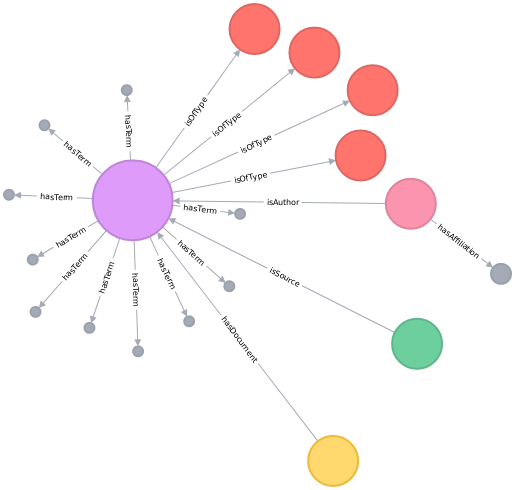}
		\caption{
			This figure is an illustration of a single document within the context graph. The document node (purple) has several gray annotation nodes, four red publication type nodes, an pink author node with a gray affiliation. The source (PubMed) is annotated in a green node, the journal in a yellow node. 
		}
		\label{fig.context0}
	\end{center}
\end{figure}

Other important context is $E_{Annotation}$ which stores multiple types of annotations such as named entities or keywords, all of which come from the MeSH tree, see \cite{rogers1963medical} and \url{https://www.nlm.nih.gov/mesh/intro_trees.html}. Therefore, $E_{MeSH}\subset E_{Annotation}$ inherently contains a hierarchy and edges $R_{MeSH}$. 
The value of MeSH terms and their hierarchy for knowledge extraction was shown in several recent studies \cite{yang2018research}. 
Figure \ref{fig.context0} depicts the knowledge graph of a single document. 

\begin{figure*}[t]
	\begin{center}
		\includegraphics[width=0.8\textwidth]{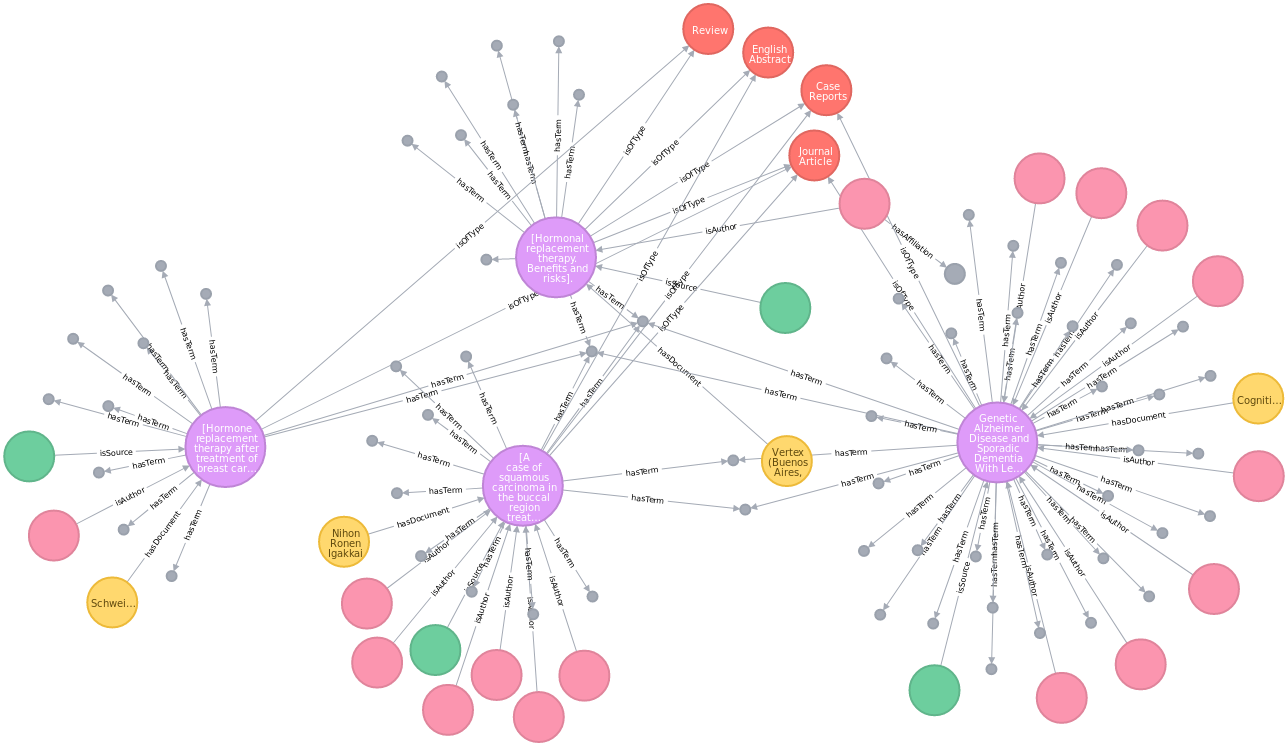}
		\caption{
			This figure is an illustration of the initial document and context graph. A PubMed node is the source of document nodes (lila). There are several context annotations like article type (red), keywords (gray), authors (pink) and journal (yellow). Authors have additional context (affiliations, gray). 
		}
		\label{fig.context1}
	\end{center}
\end{figure*}

All other relations can be added between the sets $E_{i}$, for example $R_{isCoAuthor}$, $R_{hasAffiliation}$, etc. With this information, it is -- from an algorithmic point of view -- quite easy to combine all context relations such as $R_{hasDocument}$, $R_{isAuthor}$, $R_{hasAnnotation}$, $R_{hasCitation}$ etc, though these edges should also store additional provenance information as shown in Figure \ref{fig.context1}.
  
\subsubsection{Extending the knowledge graph using NLP-technologies}

The initial knowledge graph can be extended by NLP-technologies. Terminologies and Ontologies are a widely considered topic in research during the last years. They play an important role in data and text mining as well as knowledge representation in the semantic web. They have become increasingly more important once data providers began publishing their data in a semantic web formats, namely RDF
(\cite{RDF}) and OWL (\cite{OWL}), to increase integratability.  The term
\emph{terminology} refers to the SKOS meta-model \cite{SKOS} which can be
summarized as concepts, unit of thoughts which can be identified, labeled with lexical strings, assigned notations (lexical
 codes), documented with various types of note, linked to other concepts and
 organized into informal hierarchies and association networks, aggregated, grouped into labeled and/or ordered collections, and mapped to
 concepts.  
 Several complex models have been proposed in literature and have been implemented in software, see \cite{Zeng07}. \emph{Controlled Vocabularies} contain lists of entities which may be completed to a \emph{Synonym Ring} to control synonyms. \emph{Ontologies} also present properties and can establish associative relationships which can also be done by \emph{Thesauri} or \emph{Terminologies}. See \cite{NISOZ39} and \cite{Zeng08} for a complete list of all models.   
 
 Here we define Terminologies similar to Thesauri as a set of concepts. They form a \emph{DAG} with child and parent concepts. Additionally, we have an associative relation which identifies related concepts. Each concept has at least one label, one of which is used as the preferred identifier while all others are synonyms. 
 To sum up, using ontologies or terminologies for NER has several advantages. In particular, it leads to a hierarchy within these ontologies and orders named entities according to these relations. 
 Though, we must not only consider ontologies and terminologies, but also controlled vocabularies such as MeSH. Here, we have additional annotations with different provenances, one derived as keywords with the data and one obtained from NER.

 Another example of a terminology is the Alzheimer's Disease Ontology (ADO, see \cite{MALHOTRA2014238}) $E_{ADO}$ or the Neuro-Image Terminology (NIFT, see \cite{iyappan2017neuroimaging}) $E_{NIFT}$ coming with their hierarchy $R_{ADO}$, $R_{NIFT}$. The process of NER leads to another context relation $E_{hasAnnotation}$. Since not all ontologies or terminologies are described using the RDF or OBO format, we have to add data using multiple external sources via a central tool capable of providing all the necessary ontology data. We use a semantic lookup platform containing OLS and OxO  (see \cite{madansemantic2018}).

Additional context data useful for knowledge extraction are citations such as the edges $R_{hasCitation}$ between two nodes in $E_{Document}$. Data from PMC already contains citation data with unique identifiers (PubMed IDs).
Some data is available with WikiData, see \cite{voss2016classification} and \cite{VrandecicDL2018}. Other sources are rare, but exist, see \cite{osswald2015continuing}. Especially for PubMed a lot of research is working on this difficult topic, see for example \cite{volanakis2018sciride}. 

\begin{figure}[t]
	\begin{center}
		\includegraphics[width=0.65\textwidth]{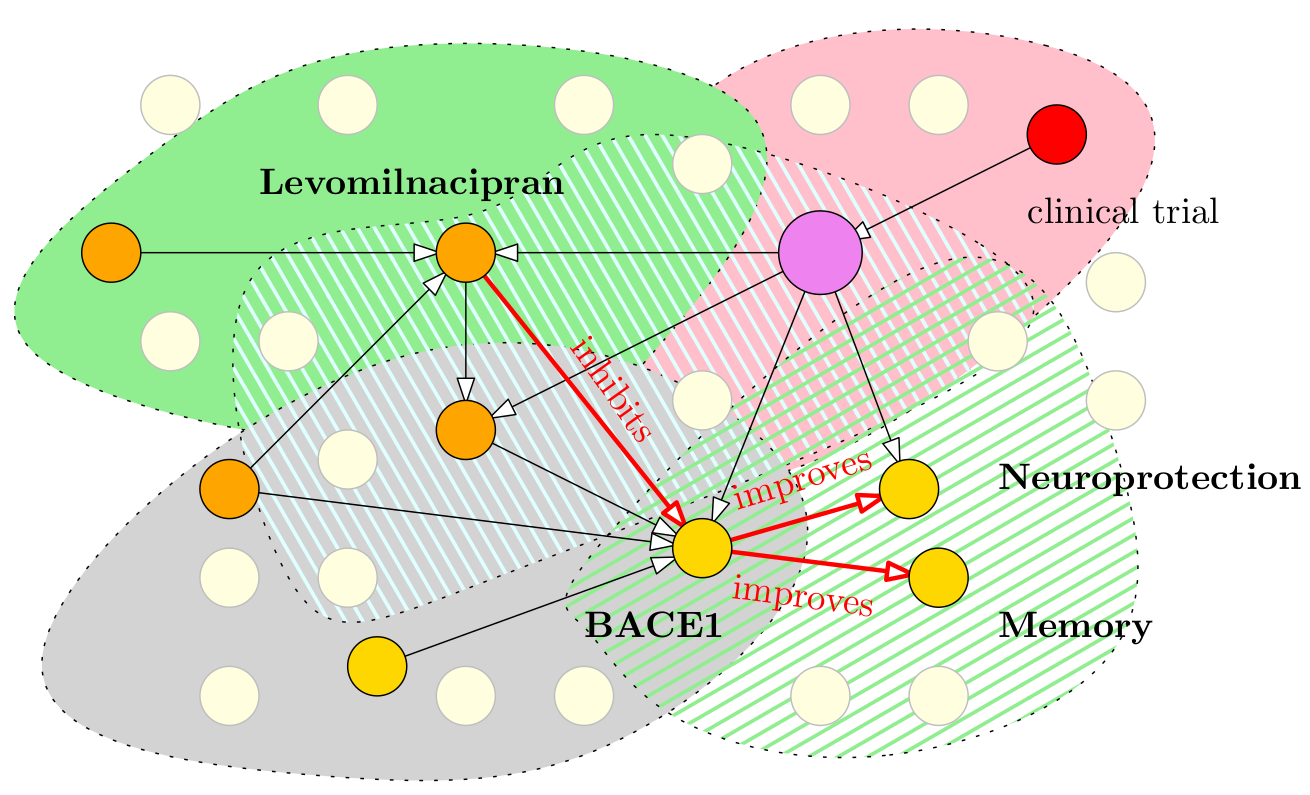}
		\caption{
			This figure is an illustration of biological knowledge within the context graph. The document node (purple) has several gray annotation nodes which come from different terminologies found with NER. The relation extraction task found the relation "Levomilnacipran" inhibts "BACE1", "BACE1" improves "Neuroprotection" and "BACE1" improves "Memory". These relations are illustrated with red edges. Since the document describes a clinical trial, this is also context for the relations as well. All other context is illustrated by colored sets, defining subgraphs. 
		}
		\label{fig.biologie}
	\end{center}
\end{figure}

\begin{figure}[t]
	\begin{center}
		\includegraphics[width=0.8\textwidth]{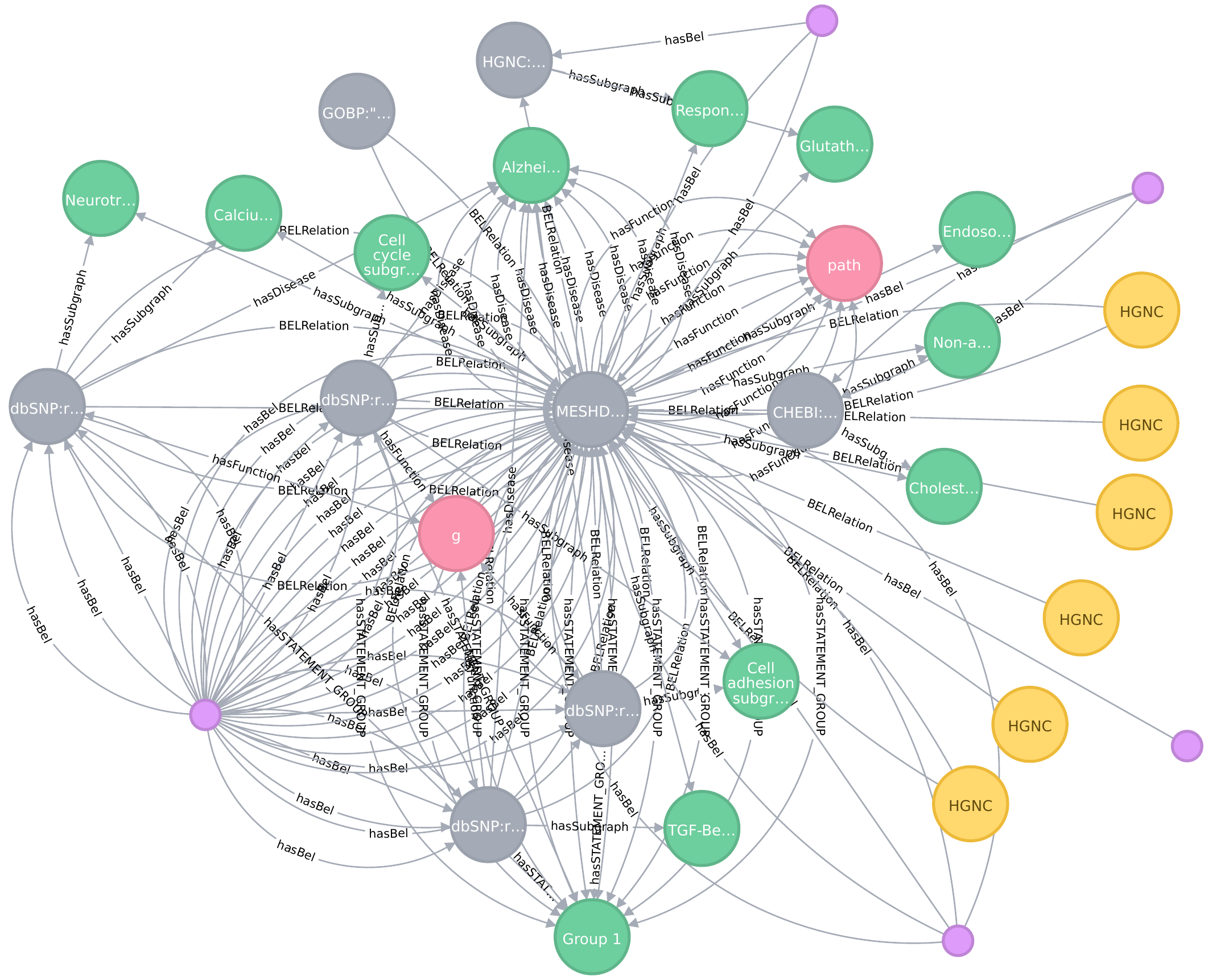}
		\caption{
			This figure is an extract  illustration of a single entity (\emph{MESHD:Alzheimers}) within the context graph. The node (gray) has several gray annotation nodes, green context nodes, documents as references (purple) and biological events (red). Whereas figure \ref{fig.biologie} shows a small example, we can see here, that the knowledge graph might get very complex. 
		}
		\label{fig.context15}
	\end{center}
\end{figure}

Furthermore, we can consider the relational information between entities. For example, BEL statements naturally form knowledge graphs by way of semantic triples that consist of concepts, functions and relationships \cite{fluck}. 
To tackle such complex tasks they constantly gather and accumulate new knowledge by performing experiments, and also studying scientific literature that includes results of further experiments performed by researchers. 
Existing solutions are primarily based on the methods of biomedical text mining which consists of extracting key information from unstructured biomedical text (such as publications, patents, and electronic health records). 
Several information systems have been introduced to support curators in generating these networks such as BELIEF, a workflow that builds BEL-like statements semi-automatically by retrieving publications from a relevant corpus generator system called SCAIView, see \cite{belief} and \cite{belief2}.

Figure \ref{fig.biologie} illustrates a few basic relations such as \emph{"Levomilnacipran" inhibts "BACE1"}, \emph{"BACE1" improves "Neuroprotection"} and \emph{"BACE1" improves "Memory"}, all of which were found using relation extraction methods on named entities in a document. It is important to note that context for a document can also be context for the derived relations and vice versa. If an entity that forms part of a relation has synonyms, or is found within another document with a different context, this may lead to a deeper understanding about the statement. An example of this interconnectedness is shown Figure \ref{fig.context15}. Due to the complexity, the resulting graph structures become difficult to manually parse and intepret thus requiring algorithmic approaches to properly analyze.


\section{Results}
\label{sec.results}

\subsection{Real world usecases for testing}

We collected 27 real world questions and queries in scientific projects. They are of varying complexity (Table \ref{tab:Fragestellungen}) and can be used to test the biomedical knowledge graph. 
Some of them use local structures, for example conjunctive regular path queries (CRPQ, see \cite{Wood:2012:QLG:2206869.2206879}) which combine subgraph pattern with queries regarding paths (problems 1,3,5,7,9,10,13,15,20) or the extended version ECRPQ (8,18,22). Other local structures include Regular Path Queries (RPQ, see \cite{Angles:2017:FMQ:3145473.3104031}) (problems 2,11,14,16,17,19,21) and finding shortest path (problems 4,12). 
Additional queries use global structures such as centrality which include Page Rank (6,23), Betweenness Centrality (25) or Degree Centrality (26). Another global problem is community detection, for example Louvain Modularity (24) or Connected Components (27).

\begin{longtable}{|p{0.3cm}|p{7.1cm}|p{2.7cm}|p{3.1cm}|}
\caption{Biomedical example queries on knowledge graphs with context data}
\tiny
\label{tab:Fragestellungen}
\endfirsthead
\endhead
\hline
\textbf{\#} & \textbf{Query} & \textbf{Input Example} & \textbf{Output}  \\ 
\hline
1 & Which author was the first to state that \{Entity1\} has an enhancing effect on \{Entity2\}? & APP, gamma Secretase Complex & Author and document title \\
\hline
2 & Which genes \{Entity1\} play a role in two diseases \{Entity2\}? & Entity.source = HGNC, MESH & subgraph of genes with 2 diseases \\
\hline
3 & In which journal was it published that \{Entity1\} has an enhancing effect on \{Entity2\}? & APP, gamma Secretase Complex & Document and Journal \\
\hline
4 & What is the shortest way between \{Entity1\} and \{Entity2\} and what is on that way? & axonal transport, LRP3 & path between nodes \\
\hline
5 & Where was it published that \{Entity1\} has an enhancing effect on \{Entity2\} and what documents cite this? & APP, gamma Secretase Complex & List of publishing and citing documents \\
\hline
6 & What are the most important entities in context of \{Entity1\} disease? & Alzheimer's & Page Rank of neighboring entities \\
\hline
7 & Which authors publish in the same journal on the topic \{Entity1\} and have not yet published together? & Alzheimer's disease & List of author couples \\
\hline
8 & Find a path of biological entities that connects \{Entity1\} with \{Entity2\} & Alzheimer's disease, ACHE & path of entities \\
\hline
9 & Are there authors within the same affiliation who make contradictory statements regarding protein \{Entity1\} and protein \{Entity2\}? & apoptotic process, SLC25A21 & number of statements for both variants \\
\hline
10 & Do the data in the literature correlate with the concomitant diseases for illness \{Entity1\}? So are the genes mentioned in \{Entity1\} documents also mentioned in \{Entity2\} documents of the concomitant disease? & Alzheimer's, Down syndrome & genes involved in both diseases in the literature \\
\hline
11 & Does the function of a gene \{Entity\} differ in different contexts? & IL1B & List of all functions in contexts \\
\hline
12 & How far apart are \{document1\} and \{document2\}? & PMID:16160056, PMID:16160050 & Shortest path between documents \\
\hline
13 & Does the biological process on gene \{Entity1\} also exist in  context of \{Entity2\}? And what author describes it? & APOE, brain & outcome graph in  context of the brain \\ 
\hline
14 & Are there BEL statements that have no source, so should be checked? & - & List of relations \\
\hline
15 & How many sources are there for the statements of a contradictory BEL statement? & hasRelation. function = increases, decreases & number of sources for each of the cases \\
\hline
16 & Is there also a relation between the documents describing the entities \{Entity1\} and \{Entity2\} that matches the relation in a BEL statement with the entities \{Entity1\} and \{Entity2\}? & APP, Alzheimer & document pairs  \\
\hline
17 & Find the oldest document describing an entity \{entity\} & APP & Oldest Document \\
\hline
18 & Is a reviewer \{Author1\} suitable for a proposal with the author \{Author\} or is there a conflict of interest? Does the reviewer have relationships with the author in the form of joint work or equal affiliation? & Ulrich Rothe, A. Castillo & Potential Graph between the authors \\ 
\hline
19 & On which topics does the author \{Author\} write most? & Ulrich Rothe & List of the most frequent annotations \\
\hline
20 & In which other journals could the author \{Author\} write with his main topics? Which journal in which he has not yet published would suit him from his main topics? & Ulrich Rothe & List of journals that could fit him \\
\hline
21 & Which Affiliation has the most publications on the topic \{Entity\} in the Journal \{Journal\}?& D008358, Biotechnology letters & Affiliation with the highest number of publications \\
\hline
22 & From when is the document cited in documents dealing with the subject \{Entity\}? & D017629 & publication date of cited document \\
\hline
23 & Which document is the most cited paper in connection with \{Entity\}, of papers that also annotate \{Entity\}? Determined by PageRank. & D017629 & Most cited paper-type document \\
\hline
24 & Which entities have many relations with \{Entity\}? Determined by Community Detection. & APP & surrounding community graph \\
\hline

25 & Which author connects the two subject areas \{Entity1\} and \{Entity2\} most strongly? & Alzheimer Disease, Parkinson & Author with highest betweenness centrality \\
\hline
26 & Which gene \{Entity\} is the most important? & Entity.source = HGNC & Entity with highest degree centrality \\
\hline
27 & Are there strongly connected components between the entities? & & Assignment of the entities to cliques \\
\hline
\end{longtable}

Because the general subgraph isomorphism problem is known to be NP-complete, we expect that some of our queries, such as finding the shortest paths in P, to require a wide range runtimes. 
The queries given in Table \ref{tab:Fragestellungen} are formulated as Cypher-queries. Query 2 is a relatively simple query given by \texttt{match (sickness1:Entity {source: "{}MESH"{}}) <-[:hasRelation]- (gene:Entity {source: "{}HGNC"{}}) -[:hasRelation]-> (sickness2:Entity {source: "{}MESH"{}}) return gene, sickness1, sickness2}, however, more complex subgraph patterns can also be generated such as Query 20 which is given by \texttt{match p=(author:Author {forename: "{}Ulrich"{}, surname: "{}Rothe"{}})-[:isAuthor]-(doc:Document)-[:isAuthor]-(reviewer:Author {forename: "{}A"{}, surname: "{}Castillo"{}}), p2=(author) -[]- (doc1:Document)-[:hasCitation]- (doc2) -[:isAuthor]- (reviewer) , p3=(author)-[]- (a:Affiliation)- []-(reviewer) return p,p2,p3 limit 10}. 

\subsection{Storing the Knowledge Graph}

Storing all of the data in one graph database without using Redis (Full) uses 58,9 GB of memory, while 
Poly1 only uses 50,82 GB (Neo4j) and 0,9 GB (Redis) of memory. The third system, Poly2, uses 50,74 + 10,2 GB (Neo4j) and 1,4 GB (Redis) memory. 

The import data is about 50 GB and generates nearly 160M nodes with relations. 
These nodes are merged by Neo4j to unique nodes. In the end we obtained 71M unique nodes and 860M relationships. 
Given the input data, we create \~{}30M nodes describing documents from PubMed and PMC, about 17M dedicated to authors, 21M affiliations and around 5M entities. The graph contains 554M annotation relationships and in total 850M relationships.

\subsection{Polyglot persistence systems}
\begin{figure}[t]
	\begin{center}
		\includegraphics[width=0.9\textwidth]{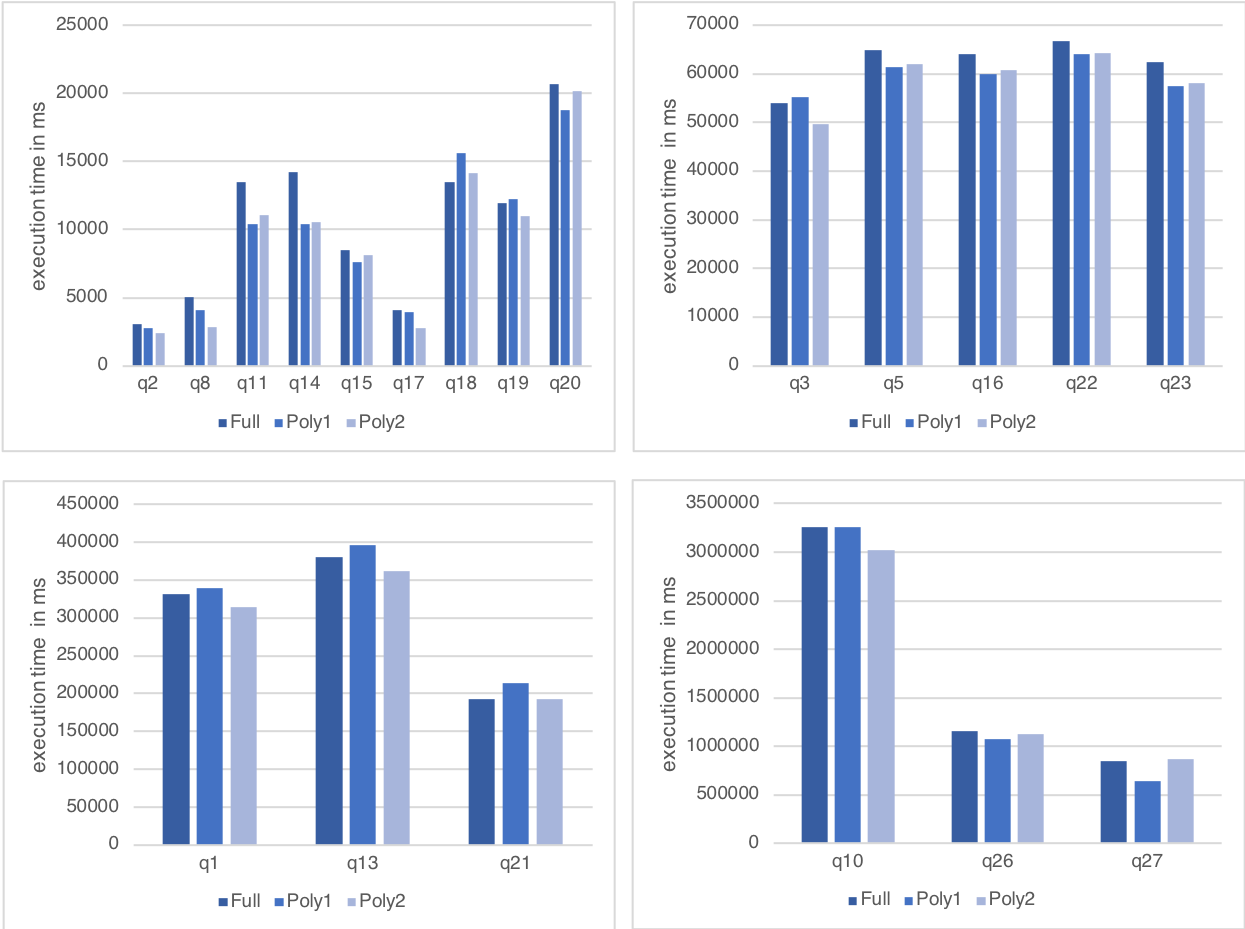}
		\caption{
			Runtime results of 27 real world queries. The queries are grouped in four diagrams with similar runtimes for a better overview. We see that the execution time of most queries is improved with \textit{Poly1} and \textit{Poly2}. In the best case the improvement is 43\%.  
		}
		\label{fig.Messungen}
	\end{center}
\end{figure}

Figure \ref{fig.Messungen} shows the runtime results of the 27 real world queries described in Table \ref{tab:Fragestellungen}.
We see that execution of some queries required a large amount of time with the longest query taking more than one hour. 
Interestingly, the execution time for most of the queries improved when ran using either the Poly1 or Poly2 implementation.  
Seven out of the 27 queries did not terminate.

For most queries, the polyglot persistence systems achieve better results, in the best case up to 43\%. However, there are differences between the systems for a few of the queries tested in that \textit{Poly1} can sometimes have better results than \textit{Poly2} and vice versa.
Contrary to expectations, \textit{Full} was found to have the best query time in most cases. 
The advantage of \textit{Poly1} over \textit{Poly2} can be explained by the fact that the memory consumption of \textit{Poly2} increased significantly due to the process of converting from string to integer and therefore the execution of the queries is slowed down. For the queries in which \textit{Poly2} performed better, this can be explained by the fact that the queries take advantage of the optimized polyglot data schema despite the higher memory consumption of the database. This is significant for example in queries 8 and 17. 

The differences in the results become clearer when you look at the differences in runtimes in percent and compare them with each other. 
The differences in the observed running times becomes clearer when analyzing the percent change in the runtime when compared to \textit{Full} as shown in Table \ref{tab:ProzentVerbesserung}.
For both systems, the average percent decrease in runtimes is calculated for all queries, in order to compare both polyglot systems each other and with \textit{Full}.   
\begin{table}
\caption{Decrease of the terms of $t_{poly1}$ and $t_{poly2}$ compared to $t_{full}$ in \%, sorted by Poly1 decreasing.}
\begin{center}
\begin{tabular}{|p{2.5cm}|p{2cm}|p{2cm}|p{3cm}|}
\hline
\textbf{Query} & \textbf{Poly1} & \textbf{Poly2} & Problem \\
\hline
14 & 26,8\% & 25,8\% & RPQ\\
\hline
27 & 23,8\% & -2,6\% & Connected Components\\
\hline
11 & 22,5\% & 17,7\% & RPQ\\
\hline
8 & 18,2\% & 43,3\% & ECRPQ\\
\hline
2 & 11,5\% & 22,9\% & RPQ\\
\hline
15 & 10,3\% & 4,5\% & CRPQ  \\
\hline
20 & 9,2\% & 2,5\% & CRPQ  \\

\hline
23 & 7,7\% & 6,8\% & Page Rank\\
\hline
26 & 6,8\% & 2,4\% & Degree Centrality \\
\hline
16 & 6,6\% & 5,1\% & RPQ\\

\hline
5 & 5,4\% & 4,6\% & CRPQ \\
\hline
22 & 3,8\% & 3,5\% & ECRPQ\\
\hline
17 & 3,1\% & 31,9\% & RPQ\\

\hline
10 & -0,2\% & 7,0\% & CRPQ  \\ 
\hline
3 & -2,3\% & 7,9\% & CRPQ \\
\hline
19 & -2,3\% & 8,0\% & RPQ\\
\hline
1 & -2,5\% & 4,9\% & CRPQ \\
\hline
13 & -4,1\% & 4,8\% & CRPQ \\
\hline
21 & -11,0\% & -0,3\%  & RPQ\\
\hline
18 & -15,7\% & -15,1\% & ECRPQ\\

\hline
\hline 
\textbf{Average} & 5,8\% & 9,8\% &\\
\hline
\end{tabular}
\label{tab:ProzentVerbesserung}
\end{center}
\end{table}
There is no information for queries 4, 6, 7, 9, 12, 24 and 25, for which no runtime could be determined on the systems as they did not go to completion. These queries are primarily graph algorithms categorized as \textit{local and global structures} in the schema discussed earlier. 

The results do not show a clear trend for any of the categories discussed. 
The \textit{RPQ} class improves on average by 15.8\% while the \textit{ECRPQ} class by 10.5\%. The classes \textit{CRPQ}, \textit{Page Rank}, \textit{Degree Centrality} and \textit{Connected Components} are in the single-digit percentage range. In general, the subcategories of \textit{local structures} seem to benefit more from the polyglot persistence designs. In addition, there is a tendency for queries that only need to consider a few node and edge types (often \texttt{entity} and \texttt{hasRelation}) to experience a greater decrease in runtimes than queries with many node and edge types.  

\subsection{Graph Queries}

Here, we present results of some of those 27 queries introduced. 
Query 1 returns a subgraph: Which author was the first to state that \{Entity1\} has an enhancing effect on \{Entity2\}?
We may execute this query using
\texttt{match (n:Entity {preferredLabel: "{}APP"{}})-[r:hasRelation {function: "{}increases"{}}]->(m:Entity {preferredLabel: "{}gamma Secretase Complex"{}}), (doc:Document {documentID: r.context})<-[r2:isAuthor]-(author:Author) return doc, author order by doc.publicationDate limit}.
\begin{figure}[t]
	\begin{center}
		\includegraphics[width=0.11\textwidth]{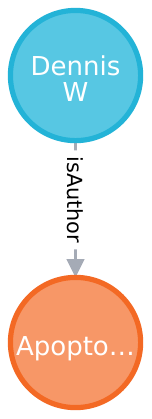} \quad\quad\quad\quad\quad\quad
		\includegraphics[width=0.45\textwidth]{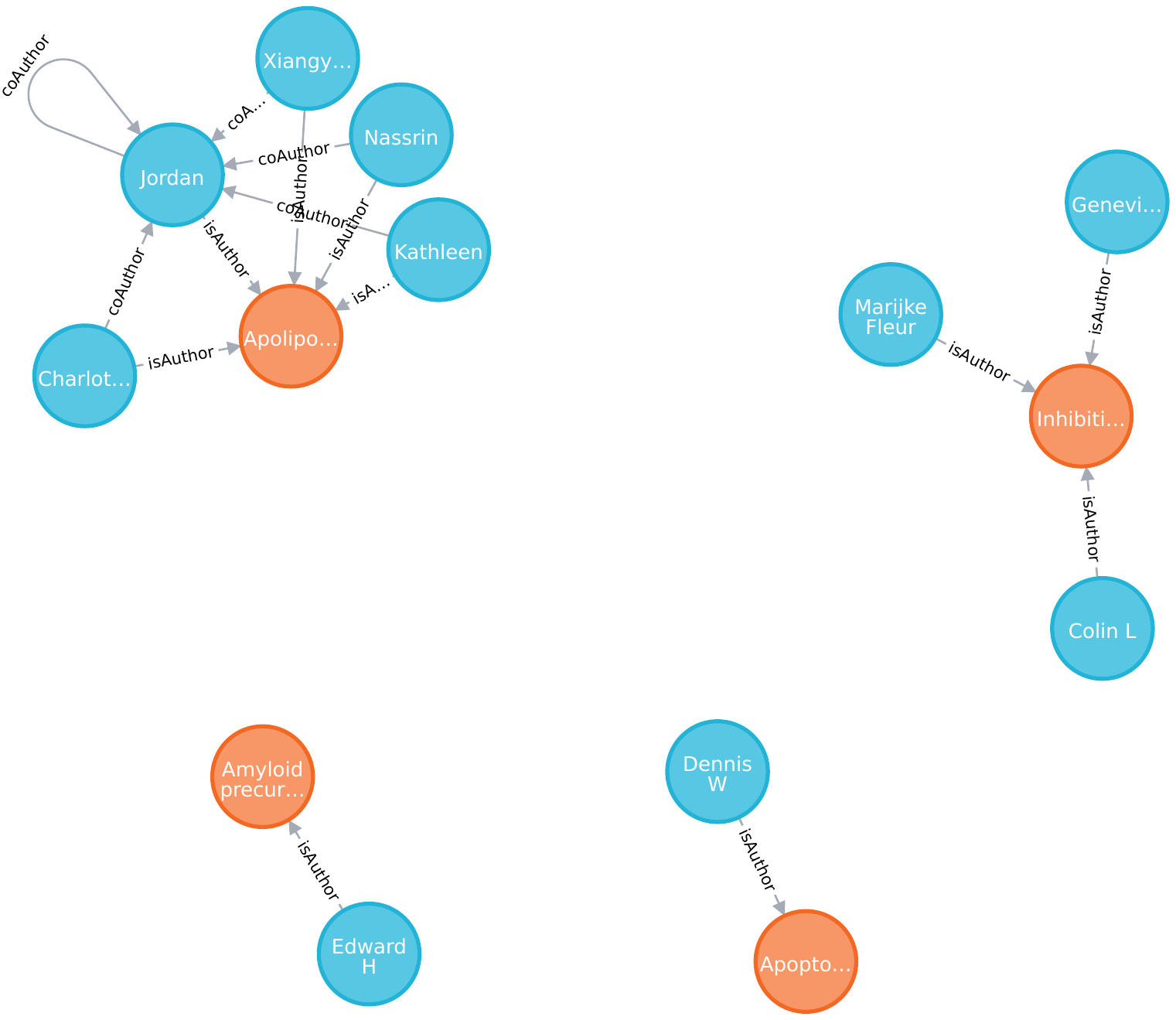}
		\caption{
			A result subgraph example for query 1: Which author was the first to state that \{Entity1\} has an enhancing effect on \{Entity2\}? On the left the first author (blue node) and the publication (orange), on the left the result shows the most recent 10 authors (blue) with their publications on this topic (orange).
		}
		\label{fig.query1}
	\end{center}
\end{figure}
A result graph can be found in figure \ref{fig.query1}. On the left the isAuthor relation with the most recent author can be found. On the left the limit parameter was changed to 10 and thus the result graph shows the most recent 10 publications and authors. 

\begin{figure}[t]
	\begin{center}
		\includegraphics[width=0.45\textwidth]{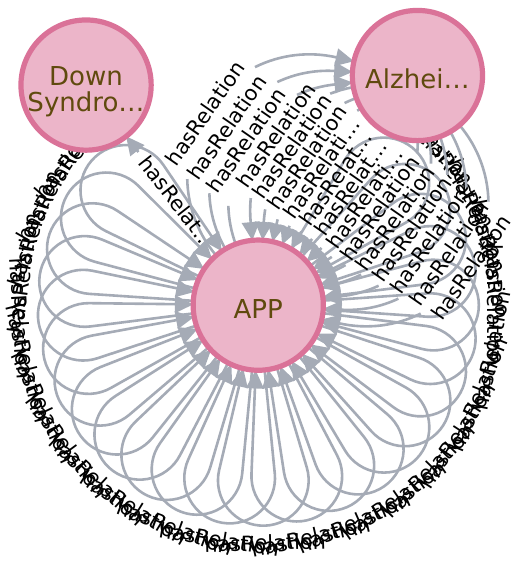}
		\caption{
			A result subgraph example for query 2: Which genes \{Entity1\} play a role in two diseases \{Entity2\}?
		}
		\label{fig.query2a}
	\end{center}
\end{figure}

\begin{figure}[t]
	\begin{center}
		\includegraphics[width=0.45\textwidth]{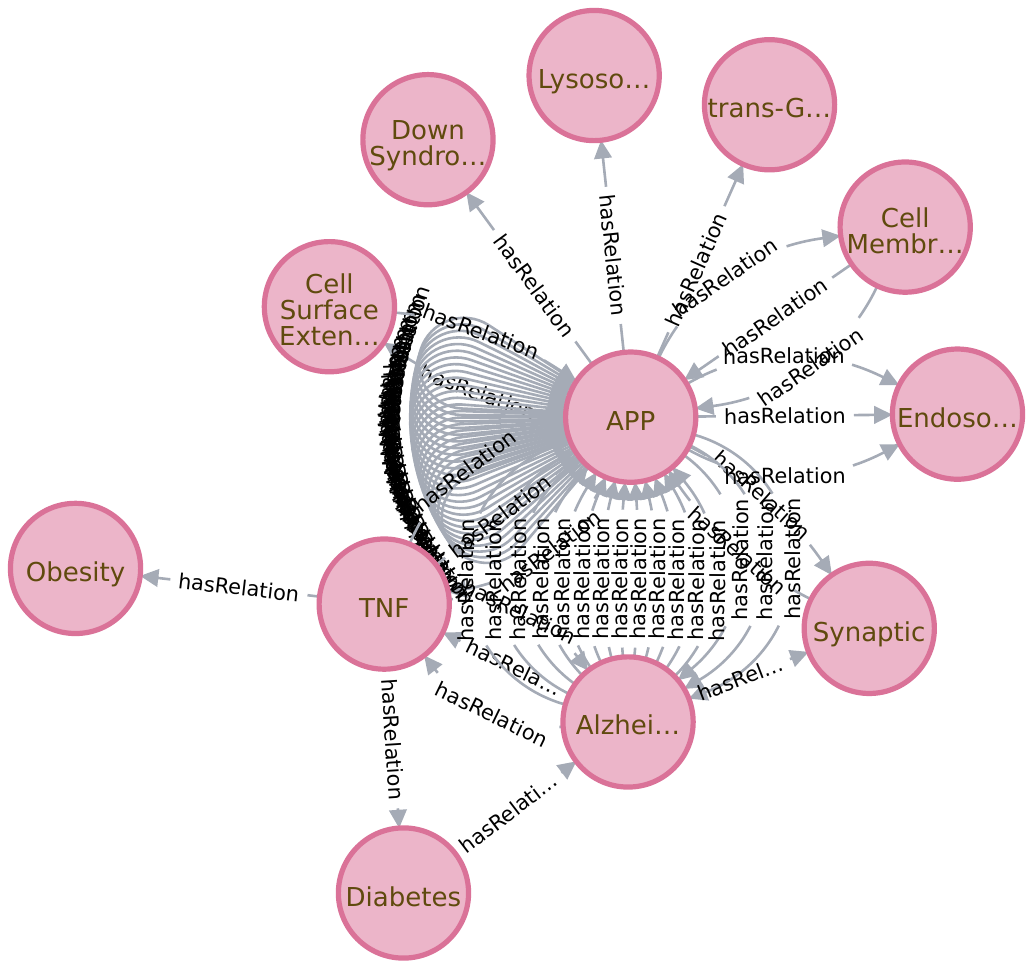}
		\caption{
			A result subgraph example for query 2 without limitation to two distinct diseases: Which genes \{Entity1\} play a role in two diseases \{Entity2\}?
		}
		\label{fig.query2b}
	\end{center}
\end{figure}

Query 2 returns a subgraph: Which genes \{Entity1\} play a role in two diseases \{Entity2\}? We may execute this query using \texttt{match (sickness1:Entity {source: "{}MESH"{}}, preferredLabel:"{}Alzheimer Disease"{}) <-[:hasRelation]- (gene:Entity {source: "{}HGNC"{}}, preferredLabel: "{}Down Syndrome"{})
-[:hasRelation]-> (sickness2:Entity {source: "{}MESH"{}}) return gene, sickness1, sickness2 limit 25}. One example output graph can be found in figure \ref{fig.query2a}.
Due to the limitation of our model to Alzheimer's Disease, it is not surprising to find only one gene -- APP. If we remove the limitation to two distinct diseases, the database returns a larger graph, see figure \ref{fig.query2b}.
Here we see, that we may need to utilize inherent ontology information to filter those nodes, that cover diseases. But we also see a second gene -- TNF -- with other diseases like Diabetes. 


Other queries return no subgraph, but rather values. For example query 25 may use built in functions from cypher:
\texttt{CALL algo.degree('MATCH (e:Entity {source: "{}HGNC"{}}) RETURN id(e) as id', 'MATCH (e1:Entity) <-[:hasAnnotation]- (d1:Document) RETURN id(e1) as source, id(d1) as target',{graph:'cypher', write:false})}.
This query answers the question "Which gene \{Entity\} is the most important?" as it returns the entity with highest degree centrality. 


\section{Discussion and Conlusion}

Here we introduce the graph-theoretic foundation for a general context concept within semantic networks and show a proof-of-concept based on biomedical literature and text mining. Our test system contains a knowledge graph derived from PubMed data which is then enriched with text mining data and domain specific language data coming from BEL. This dense graph has more than 71M nodes and 850M relationships. We discuss the impact of this novel approach using 27 real world use cases and graph queries. 

This proof-of-concept of a biomedical knowledge graph combines several sources of data by relating their contextual data to one another. We processed data from PubMed and PMC which generated more than 30M document and metadata nodes. This initial knowledge graph was extended using results from text mining and NLR-tools already included in our software as well as with named entities from ontologies also stored in SCAView. In addition, we added data generated by domain specific languages such as BEL. Thus, we were able to assess both small datasets as well as large collections of data. 

There were several issues with data integration and missing data. Initially, we tried to integrate publication data from several external sources, but some publishers used OCR technologies to convert PDF documents in XML structures. These proved problematic to process as some fields were either missing or incorrectly filled out. 

We have not yet solved the issue of author and affiliation disambiguation which remains a widely discussed topic, see \cite{kim2019correction}. An interesting novel approach -- also based on Neo4j database technology -- was introduced in \cite{franzoni2019topological}. Franzoni used topological and semantic structures within the graph for author disambiguation. Taking this into consideration, we plan to integrate such state-of-the-art technologies into our software in the future. 


Furthermore, performance for some semantic queries remains a major problem due to the massive latency for request. Although the software is integrating in our microservice architecture, see \cite{scaiview}, some queries did not run to completion. 
Here we attempt to improve our initial setup by establishing a polyglot persistence architecture in the database backend \cite{dorpinghaus2019knowledge}.
The results generated through this modification are very encouraging and we will discuss additional topics for further research.

Storing and querying a giant knowledge graph as a labeled property graph is still a technological challenge. 
Here we demonstrate how our data model is able to support the understanding and interpretation of biomedical data. 
We present several real world use cases that utilize our massive, generated knowledge graph derived from PubMed data and enriched with additional contextual data. 
Finally, we show a working example in  context of biologically relevant information using SCAIView.


\section*{Appendix}

\subsection*{Funding}

Fraunhofer Society under the MAVO Projec.

\subsection*{Author's contributions}
   This new approach goes back to an initial idea of JD and was developed by JD, AS and BS. The datasets for evaluation were produced by MJ. The manuscript was written by JD, AS and BS. All authors read and approved the final manuscript.

\subsection*{Acknowledgements}
   Valuable suggestions during the development of this method were provided by J\"urgen Klein and Vanessa Lage-Rupprecht. 
We thank Tim Steinbach for providing some illustrations to this work. In addition we thank Alexander Esser for his input on the initial research paper.
We thank Martin Hofmann-Apitius for supporting this research activity and his valuable input. 


\bibliographystyle{bmc-mathphys} 
\bibliography{lit}      







%
%

\end{document}